\documentclass[journal,letterpaper,twoside,twocolumn]{IEEEtran}
\usepackage{amsmath,amssymb,amscd,latexsym,dsfont,mathtools}
\usepackage{float}
\usepackage{color}
\usepackage{graphicx}
\usepackage{comment}
\usepackage{subfigure}
\usepackage{enumerate}
\usepackage{stfloats}
\usepackage{psfrag}
\usepackage{setspace}
\usepackage{cite}
\usepackage{balance}
\usepackage{multirow}
\usepackage[para]{threeparttable}
\usepackage{pstricks}

\renewcommand{\markboth}[1]{\renewcommand{\leftmark}{#1}\renewcommand{\rightmark}{#1}}
\newcommand{\qfunc}[0]{\mathrm{Q}}
\newcommand{\refl}[0]{\mathrm{refl}}
\newcommand{\inv}[0]{\mathrm{neg}}
\newcommand{\trans}{\mathsf{T}}

\newcommand{\SNR}[0]{{\gamma}}
\newcommand{\Es}{E_{\mathrm{s}}}

\newtheorem{theorem}{Theorem}
\newtheorem{example}{Example}

\newtheorem{remark}{Remark}

\newcommand{\PC}{P_{\C}}
\newcommand{\PtC}{\tilde{P}_{\C}}
\newcommand{\Pt}{\tilde{P}}

\newcommand{\Pj}{P_{j}}

\newcommand{\C}{\mathbb{C}}
\newcommand{\X}{\mathbb{X}}

\newcommand{\xx}{\boldsymbol{x}}
\newcommand{\bb}{\boldsymbol{b}}
\newcommand{\cc}{\boldsymbol{c}}

\newcommand{\pp}{\boldsymbol{p}}
\newcommand{\qq}{\boldsymbol{q}}
\newcommand{\aalpha}{\boldsymbol{\alpha}}
\renewcommand{\aa}{\boldsymbol{a}}
\newcommand{\setX}{\mathcal{X}}
\newcommand{\setR}{\mathcal{R}}
\newcommand{\setW}{\mathcal{W}}

\newcommand{\setK}{\mathcal{K}}
\newcommand{\tab}{\,\,\,}
\newcommand{\tabb}{\,\,}
\newcommand{\betat}{\tilde{\beta}}

\newcommand{\eqsref}[2]{(\ref{#1})--(\ref{#2})}
\newcommand{\figref}[1]{Fig.~\ref{#1}}

\newcommand{\tabref}[1]{Table~\ref{#1}}

\newcommand{\theref}[1]{Theorem~\ref{#1}}
\newcommand{\thesref}[2]{Theorems~\ref{#1} and \ref{#2}}
\newcommand{\secref}[1]{Sec.~\ref{#1}}

\newcommand{\exref}[1]{Example~\ref{#1}}

\title{On the Exact BER of Bit-Wise Demodulators for One-Dimensional Constellations}
\author{%
\IEEEauthorblockN{Mikhail Ivanov, Fredrik Br\"{a}nnstr\"{o}m, Alex Alvarado, Erik Agrell\\}
\thanks{Research supported by the Swedish Research Council, Sweden (under grant \#621-2006-4872 and \#621-2011-5950) and by the European Community's Seventh's Framework Programme (FP7/2007-2013) under grant agreement No. 271986.

M. Ivanov, F. Br\"{a}nnstr\"{o}m, and E.~Agrell are with the Dept.~of Signals and Systems, Chalmers Univ.~of Technology, SE-41296 G\"oteborg, Sweden (emails: \{mikhail.ivanov,fredrik.brannstrom,agrell\}@chalmers.se). A.~Alvarado is with the Dept.~of Engineering, University of Cambridge, Cambridge CB2 1PZ, United Kingdom (email: alex.alvarado@ieee.org).}
}

\begin{document}
\maketitle

\begin{abstract}
The optimal bit-wise demodulator for $M$-ary pulse amplitude modulation (PAM) over the additive white Gaussian noise channel is analyzed in terms of uncoded bit-error rate (BER). New closed-form BER expressions for $4$-PAM with any labeling are developed. Moreover, closed-form BER expressions for 11 out of 23 possible bit patterns for $8$-PAM are presented, which enable us to obtain the BER for $8$-PAM with some of the most popular labelings, including the binary reflected Gray code and the natural binary code. Numerical results show that, regardless of the labeling, there is no difference between the optimal demodulator and the symbol-wise demodulator for any BER of practical interest (below $0.1$).
\end{abstract}


\section{Introduction and Motivation}\label{sec:intro}

Current wireless communication systems are based on the bit-interleaved coded modulation (BICM) paradigm introduced in \cite{Zehavi92may} and later studied in \cite{Caire98,Fabregas08_Book}. One key element in these systems is the demodulator which calculates logarithmic likelihood ratios (LLR, also known as L-values) for the received bits, which are then passed to the channel decoder. The calculation of L-values is crucial in many other coded systems.  The coded performance analysis of BICM systems is generally not straightforward and is usually carried out either numerically by Monte-Carlo simulation or in terms of lower and upper bounds \cite[Sec.~4]{Caire98}, \cite[Ch.~4]{Fabregas08_Book}. In this paper, we analyze the \emph{uncoded} performance of bit-wise demodulators over the additive white Gaussian noise (AWGN) channel. 

The optimal bit-wise demodulator (BD) minimizing the BER implies the calculation of (exact) L-values for the received bits. The uncoded performance of such a demodulator has been studied in \cite{Simon05feb}, where closed-form expressions for the BER for $4$-PAM with the binary reflected Gray code (BRGC) \cite{Gray53,Agrell04dec,Agrell07} are presented. Due to the complexity of the BD, the calculation of L-values in practical systems is usually done based on the so-called max-log approximation \cite[eq.~(5)]{Viterbi98feb}, \cite[eq.~(1)]{Tsgr1_15_1093}. We call this demodulator the approximate BD (ABD). The ABD is equivalent to the symbol detector in terms of uncoded BER~\cite[Sec.~IV-A]{Fertl12}, whose performance is well documented in literature, e.g., \cite[Ch.~5]{Proakis00_Book}, \cite[Ch.~10]{Simon95_Book}, \cite{ChoYoon02jul,Lee86may,Lassing03nov,Agrell04dec,Lassing03b,Szczecinski06b,LiZhang08} and references therein. 


It is well known that the uncoded BER of one-dimensional constellation can be expressed as a sum of Gaussian Q-functions, cf.~\cite[Ch.~5]{Proakis00_Book}, \cite[Ch.~10]{Simon95_Book} and references therein. The arguments of the Q-functions depend on the points that separate the decision regions associated to different bits. We refer to these points as thresholds. The computation of the thresholds for the BD---the optimal bit-wise demodulator---is in general complicated and unknown. In this paper, however, we show that this problem can be solved analytically for $4$-PAM and any labeling, extending the results presented in \cite{Simon05feb}. Moreover, we also analytically calculate the thresholds for $8$-PAM with some relevant labelings, including the BRGC, the natural binary labeling (NBC) \cite[Sec.~II-B]{Agrell11}, the folded binary code (FBC) \cite{Lassing03b} \cite[Sec.~II-B]{Agrell11}, the binary semi-Gray code (BSGC) \cite[Sec.~II-B]{Agrell11}, and the anti-Gray code (AGC)~\cite{Brink98b}. Numerical results show that optimal and suboptimal demodulators are different in terms of the BER only at a very low SNR. At BER below $0.1$ there is no notable difference between them.

The rest of the paper is organized as follows. In \secref{sec:syst_mod} we introduce the notation convention, the system model, and the two demodulators. In \secref{sec:BER_general} the BER analysis is presented. The patterns that form a labeling are studied in \secref{sec:patterns}. The threshold computation for the BD is shown in \secref{sec:Thresholds} and the numerical results in \secref{sec:numerical_results}. The conclusions are drawn in \secref{sec:conclusions}.

\section{Preliminaries}\label{sec:syst_mod}

\subsection{Notation Convention}

In this paper the following notation is used. Lowercase letters $x$ denote real or complex scalars and boldface letters $\xx$ denote a row vector of scalars. The complex conjugate of $x$ is denoted by $x^*$. Blackboard bold letters $\X$ denote matrices with elements $x_{i,j}$ in the $i$th row and the $j$th column and $(\cdot)^\trans$ denotes transposition. Calligraphic capital letters $\setX$ denote sets, where the set of real numbers is denoted by $\setR$. The  binary complement of $x \in \{0, 1\}$ is denoted by $\bar{x}=1-x$ and its bipolar representation by $\check{x} = 2x-1$. Binary addition (exclusive-OR) of two bits $a$ and $b$ is denoted by $a\oplus b$. Random variables are denoted by capital letters $X$ and probabilities by $\Pr\{\cdot\}$. 
The Gaussian Q-function is defined as $\qfunc(x) \triangleq \left(1/\sqrt{2\pi}\right)\int_{x}^{\infty}\exp(-t^2/2)\,\mathrm{d}t$.

\subsection{System Model}

In this paper we analyze a system where a vector of binary data $\bb = [b_1,\dots,b_m]$ is fed to a modulator. The modulator carries out a one-to-one mapping from $\bb$ to one of the $M$ constellation points $x \in \setX = \{s_1,\dots, s_M\}$ for transmission over the physical channel, where $M = 2^m$. We assume that $s_1<s_2<\ldots<s_M$. The modulator is defined as the function $\Phi: \{0, 1\}^m \rightarrow \setX$.

The modulator is defined by the constellation and its binary labeling. A binary labeling is specified by the matrix $\C = [\cc_1^\trans, \dots, \cc_M^\trans]^\trans$ of dimensions $M$ by $m$, where the $i$th row $\cc_i = [c_{i,1},\dots, c_{i,m}]$ is the binary label of the constellation point $s_i$, i.e., $\Phi(\cc_i)=s_i$.

For PAM constellations, $s_i = -d(M-2i+1), i=1,\dots,M$, where $d = \sqrt{{3}/{(M^2-1)}}$ to normalize the constellation to unit average energy, i.e., $\Es = (1/M)\sum_{i=1}^M{s_i^2} = 1$. We assume the bits to be independent and identically distributed (i.i.d.) with $\Pr\{B_j = u\} = 0.5$,\,$\forall j$ and $u \in \{0,1\}$, and thus, the symbols are equiprobable, i.e., $\Pr\{X = s_i\} = 1/M$,\, $\forall i$.

In this paper we consider a discrete time memoryless AWGN channel with output $y = x + \eta$, where $x\in \setX$ and the noise sample $\eta$ is a zero-mean Gaussian random variable with variance $N_0/2$. The conditional probability density function (PDF) of the channel output given channel input is
\begin{equation}
    p_{Y|X}(y|x) = \sqrt{\frac{\SNR}{\pi}}\mathrm{e}^{-\SNR(y - x)^2},
    \label{eq:gauss}
\end{equation}
where the average signal to noise ratio (SNR) is defined as $\SNR \triangleq {\Es}/{N_0} = {1}/{N_0}$.

The observation $y$ is used by the demodulator to decide on the received binary sequence, i.e., to produce $\hat{\bb} = [\hat{b}_1, \dots, \hat{b}_m]$. In this paper we consider two demodulators to obtain $\hat{\bb}$ from $y$, which are described in the next section.

\subsection{Demodulators}

The BD calculates (a posteriori) L-values for the $m$ bits based on the observation $y$, i.e.,
\begin{align}
    l_j(y) &\triangleq \log{\frac{\Pr\{B_j = 1|Y=y\}}{\Pr\{B_j = 0|Y=y\}}} \label{eq:L-value.0}\\
    &= \log{ \frac{\sum_{x \in \setX_{j, 1}}{\mathrm{e}^{-\SNR(y-x)^2}}}{\sum_{x \in \setX_{j, 0}}{\mathrm{e}^{-\SNR(y-x)^2}}}},
    \label{eq:L-value}
\end{align}
for $j=1,\dots,m$ and $\setX_{j,u} \triangleq \{s_i \in \setX: c_{i,j} = u, \forall i \}$. To pass from \eqref{eq:L-value.0} to \eqref{eq:L-value} Bayes' rule was used together with the i.i.d.~assumption of the bits and the conditional PDF in \eqref{eq:gauss}. The BD uses the L-values in~\eqref{eq:L-value} to make a decision on the received bit according to the rule
\begin{equation}
    \hat{b}_j^{\mathrm{BD}} =
    \begin{cases}
    1 &\text{if } l_j(y) \ge 0,\\
    0 & \text{otherwise}.
    \end{cases}
\label{eq:decision_rule}
\end{equation}

The implementation of the BD in its exact form~\eqref{eq:L-value} is complicated, especially for large constellations, as it requires calculation of the logarithm of a sum of exponentials. To overcome this problem, approximations are usually used in practice. The most common approximation is the so-called max-log approximation ($\log{\sum_i \mathrm{e}^{\lambda_i}} \approx \max_i{\lambda_i}$) \cite[eq.~(3.2)]{Zehavi92may}, \cite[eq.~(9)]{Caire98}, \cite[eq.~(5)]{Viterbi98feb}, \cite[eq.~(8)]{Robertson95jun}, which used in~\eqref{eq:L-value} gives
\begin{equation}
    \tilde{l}_j(y) = \SNR\left[\min_{x \in \setX_{j, 0}}{(y-x)^2} - \min_{x \in \setX_{j, 1}}{(y-x)^2}\right].
    \label{eq:max_log_rrl}
\end{equation}
The use of the max-log approximation transforms the nonlinear relationship \eqref{eq:L-value} into a piece-wise linear relationship \eqref{eq:max_log_rrl}, as previously shown in e.g., \cite[Fig.~3]{Hyun05}, \cite[eqs.~(11)--(14)]{Raju06}.

The ABD is defined as the demodulator that applies the same decision rule~\eqref{eq:decision_rule} based on L-values calculated by~\eqref{eq:max_log_rrl}. As mentioned in~\cite[Sec.~IV-A]{Fertl12} the ABD is equivalent to the symbol detector in terms of uncoded BER.
\section{BER for One-Dimensional Constellations}\label{sec:BER_general}

The BER for a given labeling $\C$ can be expressed as
\begin{equation}
\PC 	= \frac{1}{m}\sum_{j=1}^m \Pj, \label{eq:P_C1.0}
\end{equation}
where using the law of total probability, the BER for the $j$th bit position $\Pj \triangleq\Pr\{\hat{B}_j \neq b_j|B_j=b_j\}$ can be written as
\begin{equation}
P_j = \frac{1}{M}\sum_{i = 1}^{M}{\Pr\{\hat{B}_j \neq c_{i,j}  | X = s_i\}}.
\label{eq:P_C1}
\end{equation}
The BER for the $j$th bit position $P_j$ depends only on the subconstellations $\setX_{j,0}$ and $\setX_{j,1}$ (cf.~\eqref{eq:L-value}--\eqref{eq:max_log_rrl}), i.e., $P_j$ is a function of the $j$th column of $\C$.

We define a bit pattern (or simply pattern) as a length-$M$ binary vector $\pp = [p_1,\dots,p_M] \in \{0,1\}^M$ with Hamming weight $M/2$. The labeling $\C$ can now be defined by $m$ patterns, each corresponding to one column of $\C$. We index the patterns as $\pp_w$ with $w$ being the decimal representation of the vector $\pp$, i.e., $w = \sum_{i = 1}^{M}{2^{M-i}p_i}$. For example, for $M=4$, the pattern $[0,1,0,1]$ is indexed as $\pp_5$. The BER for the labeling $\C$ does not depend on the order of its columns, and thus, the BER for the labeling $\C$ is fully determined by a set of $m$ patterns (indices) $\setW= \{w_1,\dots,w_m\}$. Based on the previous discussion, from now on we concentrate our analysis only on patterns (and not on labelings). 

To analyze the BER of a pattern (PBER), the observation space $\setR$ is split into two disjoint decision regions, i.e., $\mathcal{Y}_{0} = \{y \in \setR: \hat{b} = 0\}$ and $\mathcal{Y}_{1} = \{y \in \setR: \hat{b} = 1\}$ such that $\mathcal{Y}_{0} \cup \mathcal{Y}_{1} = \setR$. Using the definition of $\mathcal{Y}_{0}$ and $\mathcal{Y}_{1}$, the PBER for the pattern $\pp = [p_1,\dots,p_M]$ can be rewritten as
\begin{equation}
P = {\frac{1}{M}\sum_{i = 1}^{M}{\Pr\{Y \in \mathcal{Y}_{\bar{p}_{i}} | X = s_i\}}}.
\label{eq:P_j}
\end{equation}
By expressing $P$ as in \eqref{eq:P_j}, it is clear that the PBER in \eqref{eq:P_C1} can be calculated using the decision region $\mathcal{Y}_{0}$ only, as opposed to alternative approaches where \eqref{eq:P_j} is expressed in terms of the PDF of the L-values \cite[eq.~(19)]{Benjillali07}.

\subsection{Decision Thresholds}\label{sec:thresholds_intro}
\begin{figure*}[t]
\newcommand{\scale}{0.90}
\begin{center}
    \psfrag{SNR}[tc][tc]{$y$}
    \psfrag{BER}[bc][Bc][\scale]{L-value}
    \psfrag{x1}[Bc][Bc]{$s_1$}
    \psfrag{x2}[Bc][Bc]{$s_2$}
    \psfrag{x3}[Bc][Bc]{$s_3$}
    \psfrag{x4}[Bc][Bc]{$s_4$}
    \psfrag{x5}[Bc][Bc]{$s_5$}
    \psfrag{x6}[Bc][Bc]{$s_6$}
    \psfrag{x7}[Bc][Bc]{$s_7$}
    \psfrag{x8}[Bc][Bc]{$s_8$}
    \psfrag{b1}[Bc][Bc][\scale]{$\beta_1$}
    \psfrag{b2}[Bc][Bc][\scale]{$\beta_2$}
    \psfrag{b3}[Bc][Bc][\scale]{$\beta_3$}
    \psfrag{b4}[Bl][Bl][\scale]{$\beta_4$}
    \psfrag{b5}[Bl][Bl][\scale]{$\beta_5$}
    \psfrag{b6}[Bl][Bl][\scale]{$\beta_6$}
    \psfrag{b7}[Bl][Bl][\scale]{$\beta_7$}
    \subfigure[$\rho = 5.5$~dB]{\includegraphics[width=0.95\columnwidth]{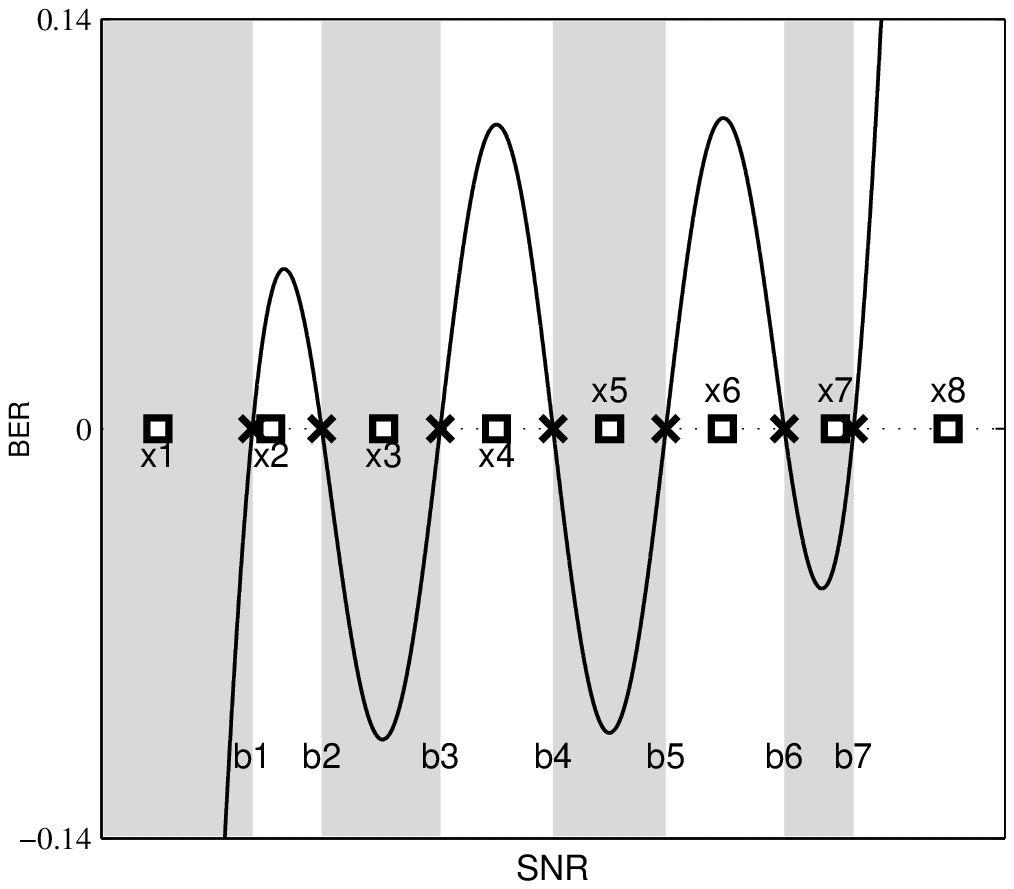}}
    \subfigure[$\rho \approx 4.9$~dB]{\includegraphics[width=0.95\columnwidth]{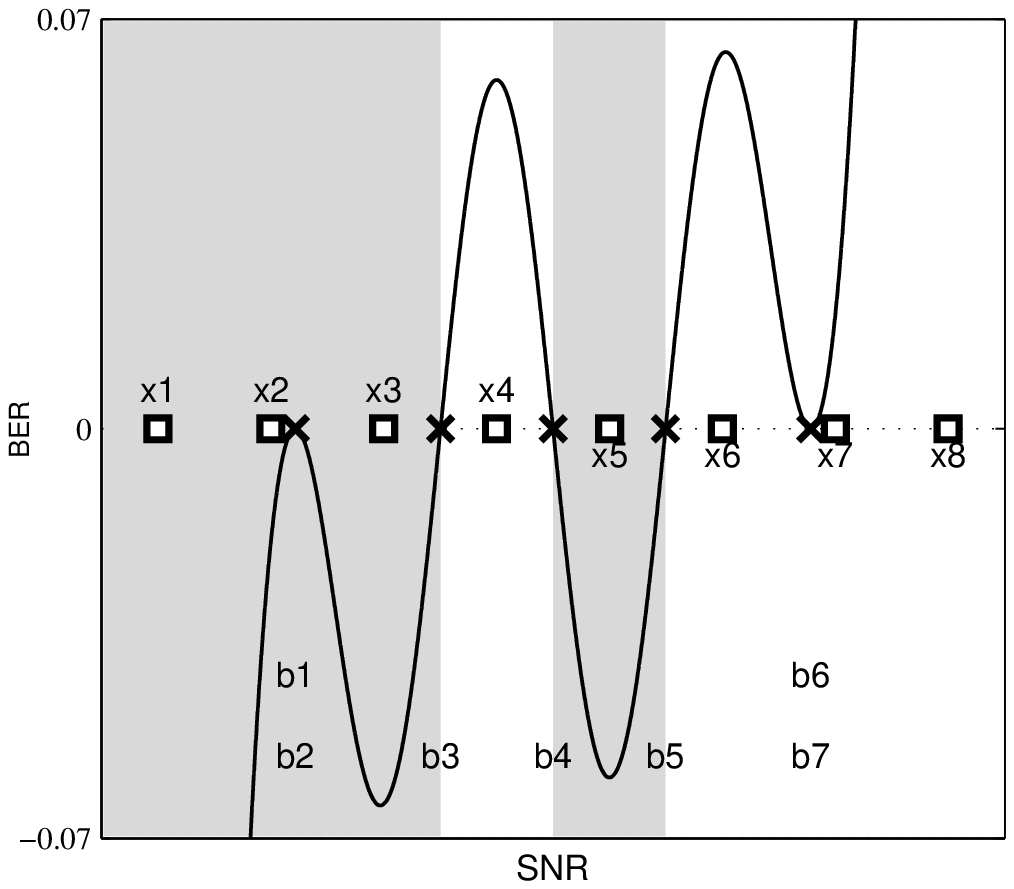}}
    \subfigure[$\rho = 4$~dB]{\includegraphics[width=0.95\columnwidth]{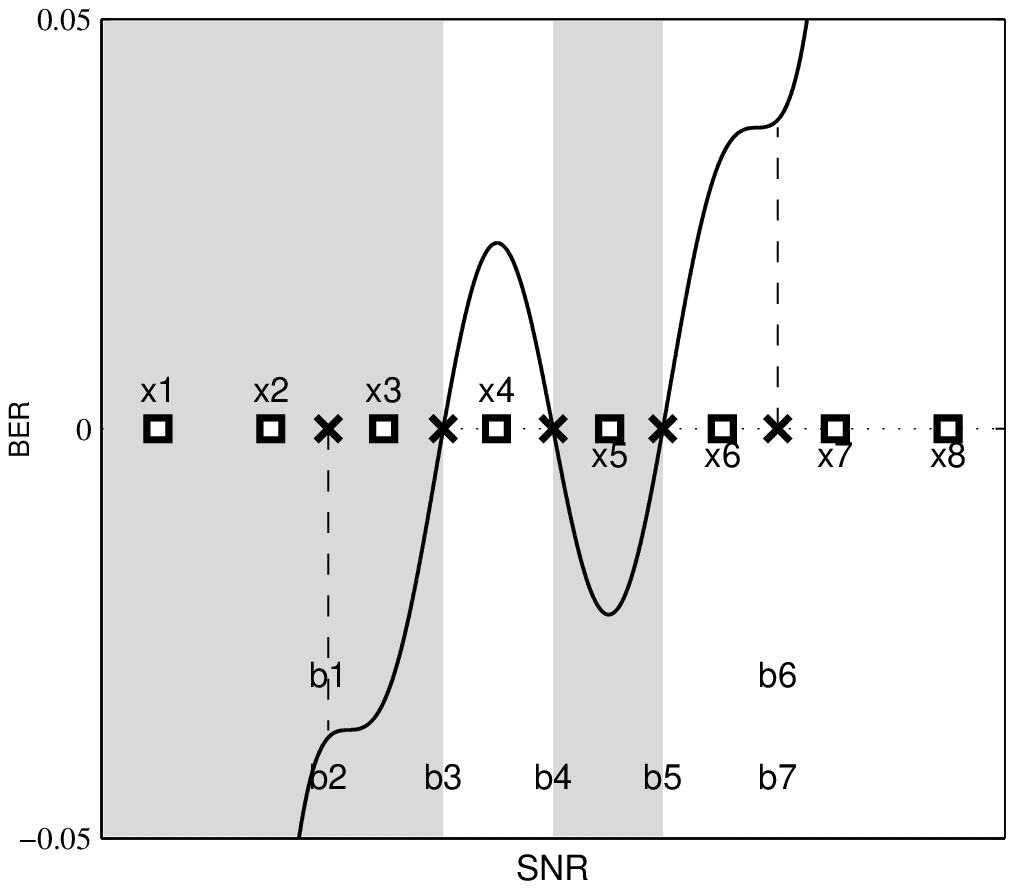}}
    \subfigure[$\rho \approx 2.2$~dB]{\includegraphics[width=0.95\columnwidth]{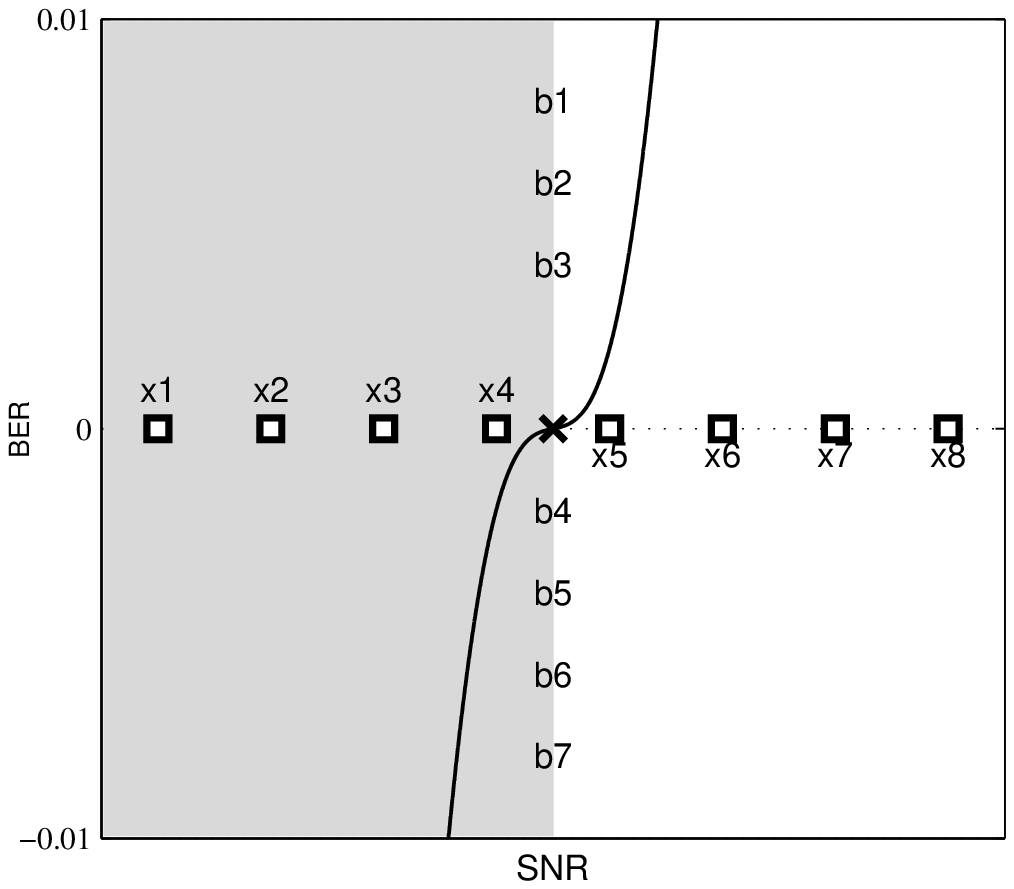}}
    \caption{L-values in~\eqref{eq:L-value} vs. the received signal for different $\SNR$ for $8$-PAM and the BD for $\pp_{85}=[0,1,0,1,0,1,0,1]$. Squares show the constellation points and crosses show the  thresholds $\beta_k$ in~\theref{theor:8pam_thresh}. Gray and white areas indicate $\mathcal{Y}_{0}$ and $\mathcal{Y}_{1}$, resp. In~(a) at $\SNR = 5.5$~dB none of the seven thresholds is virtual. At $\SNR \approx 4.9$~dB, shown in~(b), the thresholds $\beta_1$ and $\beta_2$ as well as $\beta_6$ and $\beta_7$ become virtual, shown also in~(c) for $\rho=4$~dB. At $\SNR \approx 2.2$~dB, shown in~(d), $\beta_3$ and $\beta_5$ merge with $\beta_4$ and also become virtual.}
    \label{fig:lvalues8PAM}
\end{center}
\end{figure*}

One key element in the BER analysis presented in this paper is the \emph{decision thresholds}. Decision thresholds (or simply thresholds) for a given pattern $\pp$ are the points that separate the decision regions for zeros and ones, and thus, they determine the PBER for the BD/ABD in~\eqref{eq:P_j}. For a given pattern $\pp$, we associate the threshold $\beta_k\in\mathcal{R}$ to the bit $p_k$ when $p_k\neq p_{k+1}$. Since there is no threshold $\beta_k$ when $p_k = p_{k+1}$, the number of thresholds is at most $M-1$. The indices of the thresholds for the pattern $\pp$ form a set of indices $\setK$, with $1\leq|\setK|\leq M-1$. For example, the pattern $\pp_{54}=[0, 0, 1, 1, 0 , 1, 1, 0]$ has $\setK=\{2,4,5,7\}$.

The thresholds for the ABD, which we denote by $\betat$, are independent of $\SNR$ and placed at the midpoints between adjacent constellation points with different binary labels, which follows directly from~\eqref{eq:max_log_rrl}. On the other hand, the thresholds for the BD depend on $\SNR$. We denote these thresholds by $\beta$, where to simplify the notation, the dependency on $\SNR$ is omitted. \figref{fig:lvalues8PAM} shows the thresholds for the BD and their dependency on the SNR. This figure shows that some thresholds can merge at low SNR and seem to disappear. To take this effect into account, we define virtual thresholds as follows. A threshold $\beta_k$ is said to be \emph{virtual} at $\SNR < \SNR_0$ if it merges with another threshold $\beta_{k'}$ at $\SNR = \SNR_0$ (i.e., $\beta_k=\beta_{k'}$ when $\SNR = \SNR_0$) and does not exist at $\SNR < \SNR_0$. 

\subsection{General Expression for One-dimensional Constellations}\label{sec:BER_BD}
The BER expression for the ABD and an $M$-PAM constellation with any labeling is well known and can be found in~\cite[eq.~(21)]{Agrell07}. The PBER expression can easily be obtained in a similar way. The following theorem gives a general PBER expression if none of the thresholds $\beta_k$ is virtual. 

\begin{theorem}\label{theo:ber_1D_general}
The PBER of the BD or the ABD using an arbitrary one-dimensional constellation with a pattern $\pp$ can be expressed as
\begin{align}
    P &= \frac{1}{2} + \frac{1}{M}\sum_{i = 1}^{M}\sum_{k \in \setK}g_{i,k}\qfunc\left((\beta_k-s_i)\sqrt{2\SNR}\right),\label{eq:BER_general}
\end{align}
where none of $\beta_k$ is virtual, and $g_{i,k} \in \{\pm1\}$ is
\begin{equation}
g_{i,k} \triangleq (p_{k+1}-p_k)(1-2p_i).\label{eq:g_ik}
\end{equation}
\end{theorem}

\begin{IEEEproof}
The proof is given in Appendix~\ref{Appendix.theo:ber_1D_general}.
\end{IEEEproof}

The following theorem shows that \theref{theo:ber_1D_general} also holds when some of the thresholds become virtual, provided that their values are chosen properly.

\begin{theorem}\label{theo:ber_1D_general.virtual}
If $\beta_{k}$ is virtual for $\SNR<\SNR_0$ because $\beta_k=\beta_{k'}$ at $\SNR = \SNR_0$ for some $k$ and $k'>k$, the PBER for the BD is given by~\eqref{eq:BER_general}--\eqref{eq:g_ik} if $\beta_k=\beta_{k'}, \tab \forall \SNR<\SNR_0$.
\end{theorem}
\begin{IEEEproof}
We will show that the PBER in \eqref{eq:BER_general} is not affected by the virtual threshold $\beta_k$ if $\beta_{k} = \beta_{k'}$. Let $S_i$ be the two terms in the inner sum in \eqref{eq:BER_general} associated to the thresholds $\beta_{k}$ and $\beta_{k'}$, i.e.,
\begin{multline}\label{eq:sigma}
    S_i \triangleq g_{i,k}\qfunc{\left( (\beta_{k} - s_i)\sqrt{2\SNR}  \right)} + g_{i,k'}\qfunc{\left( (\beta_{k'} - s_i)\sqrt{2\SNR}\right)}.
\end{multline}
Since $\beta_{k}$ and $\beta_{k'}$ are  thresholds that merge, $p_{k+1} = p_{k'}$ and $p_{k} = p_{k'+1}$ must hold. Using these relations in \eqref{eq:g_ik}, we obtain $g_{i,k} = -g_{i,k'}$, which used in \eqref{eq:sigma} gives $S_i=0, \tab \forall i$  if $\beta_{k} = \beta_{k'}$. \end{IEEEproof}

\begin{remark}
\theref{theo:ber_1D_general.virtual} holds regardless of whether $\beta_{k'}$ is virtual or not. If $\beta_{k'}$ is not virtual for $\SNR<\SNR_0$, $\beta_k$ must be set to the value of $\beta_{k'}$ for $\SNR<\SNR_0$. If $\beta_{k'}$ is virtual for $\SNR<\SNR_0$, $\beta_k=\beta_{k'}$ can be set to any real value.
\end{remark}

\subsection{BER for the ABD and $M$-PAM}\label{sec:BER_ABD}

For $M$-PAM and the ABD, \eqref{eq:BER_general} can be expressed as a bit-wise version of \cite[eq.~(21)]{Agrell07}:
\begin{equation}
    \Pt = \frac{1}{M}\sum_{n = 1}^{M-1}a_n\qfunc{\left( (2n-1)d\sqrt{2\SNR}\right)},
    \label{eq:BER_maxlog_new}
\end{equation}
where
\begin{multline}
    a_n = \sum_{k = n}^{M-1} (p_{k+1} - p_{k})(1-2p_{k+1-n}) \\ - (p_{k+2-n} - p_{k+1-n})(1-2p_{k+1}).
        \label{eq:ber_maxlog_weights}
\end{multline}

One direct consequence of~\eqref{eq:BER_maxlog_new} is that the vector $\aa \triangleq [a_1, \dots, a_{M-1}]$ with $a_n$ given by~\eqref{eq:ber_maxlog_weights} completely defines the performance of the ABD for $M$-PAM and allows us to compare the performance of different patterns. From~\eqref{eq:BER_maxlog_new}, the PBER for high SNR is determined by the coefficient multiplying the Q-function with the smallest argument, i.e., $a_1$. If two patterns have identical coefficients $a_1$, the next coefficients $a_2$ should be compared, and so on.

Using~\eqref{eq:P_C1.0} and \eqref{eq:BER_maxlog_new}, the average BER for an $M$-PAM with a labeling $\C$ can be expressed as
\begin{equation}\label{eq:BER_ABD_labeling}
    \PtC = \frac{1}{mM}\sum_{n = 1}^{M-1}\alpha_n\qfunc{\left( (2n-1)d\sqrt{2\SNR}\right)},
\end{equation}
where $\aalpha \triangleq [\alpha_1, \dots, \alpha_{M-1}]$ is the sum of vectors $\aa$ for the $m$ patterns used in $\C$. The equation in~\eqref{eq:BER_ABD_labeling} in fact corresponds to~\cite[eq.~(21)]{Agrell07}, where the value of $\alpha_n$ is a scaled version of the so-called differential average distance spectrum $\bar{\delta}(n,\lambda)$, i.e., $\alpha_n = 2M\bar{\delta}(n, \lambda)$.

%
%

\section{Bit Patterns}\label{sec:patterns}

For $4$-PAM, the patterns $\pp_5 = [0, 1, 0, 1]$ and $\pp_{10}= [1, 0, 1, 0]$ have identical PBER performance because of the symmetry of the constellation. To find all the patterns with different performance, we group patterns into classes, where all the patterns in the same class have identical PBER.

In the following, we define two operations that can be applied to a pattern. A \emph{reflection} of $\pp$ is defined as $\pp' = \refl{(\pp)}$ with $p'_i = p_{M +1 - i}$ for $i = 1,\dots,M$. For example, $\pp_{27} = [0,0,0,1,1,0,1,1]=\refl{([1,1,0,1,1,0,0,0])}=\refl{(\pp_{216})}$. An \emph{negation} of $\pp$ is defined as $\pp' = \inv{(\pp)}$ with $p'_i = \bar{p}_i$ for $i = 1,\dots,M$. For example, $\pp_{39} = [0,0,1,0,0,1,1,1]=\inv{([1,1,0,1,1,0,0,0])}=\inv{(\pp_{216})}$. Both these functions are self-inverse, i.e., $\pp = \refl(\refl(\pp))$ and $\pp = \inv(\inv(\pp))$, and they commute, i.e., $\refl(\inv(\pp)) = \inv(\refl(\pp))$. Note also that for any pattern $\pp$, we have that $\pp \neq \inv(\pp)$. Using the previous definitions, we now define three special types of patterns that will be useful throughout this paper.

The pattern $\pp$ is said to be \emph{reflected} (RE) if $\refl({\pp}) = \pp$. For example, $\pp_{60} = [0,0,1,1,1,1,0,0]$ is an RE pattern. The pattern $\pp$ is said to be \emph{anti-reflected} (ARE) if $\inv(\refl({\pp})) = \pp$. For example, $\pp_{43} = [0,0,1,0,1,0,1,1]$ is an ARE pattern.The pattern $\pp$ is called \emph{asymmetric} (ASY) if it is neither RE nor ARE. For example, $\pp_{216} = [1,1,0,1,1,0,0,0]$ is an ASY pattern.

From \eqsref{eq:P_C1}{eq:P_j}, we note that the PBER is not affected by reflections and/or negation of the patterns, since the PBER is averaged over both transmitted zeros and ones. Because of this,  we group all patterns that are connected via reflection or negation into one class of patterns with identical PBER. Each class contains at the least two patterns ($\pp$ and $\inv(\pp)$ because $\pp \neq \inv(\pp),\tab\forall \pp$) and at most 4 different patterns ($\pp$, $\inv(\pp)$, $\refl(\pp)$, and $\inv(\refl(\pp))$ if they are all different) and is represented by a unique class index $q\in {1,\dots, Q}$, where $Q$ is the number of classes. A labeling can now be represented not only by a binary matrix $\C$ or by the set of pattern indices $\setW$, but also by an ordered vector of class indices $\qq=[q_1,q_2,\dots,q_m]\in\{1,\dots, Q\}^m$, where $q_1\le q_2 \le \dots \le q_m$. The reason for introducing this vector $\qq$ is that it allows us to easily identify two binary labelings that give the same BER. In other words, if two different labelings $\C$ and $\C'$ have vectors $\qq$ and $\qq'$, resp., they will give the same average BER if $\qq = \qq'$. To clarify these definitions, consider the following example (see also Example~\ref{ex:8PAM}).

\begin{example}[Patterns for $4$-PAM]\label{ex:4PAM}
For $4$-PAM there are six patterns which are grouped into $Q=3$ classes as shown in \tabref{tab:4PAM_seq}. Two of them are RE and four are ARE, as indicated in the third column of~\tabref{tab:4PAM_seq}. The first column is the index of the class and the fourth column contains the decimal representations of the indices of the patterns that belong to that class. As an example, one of the patterns of a class is shown in the second column. These patterns are called \emph{representative} and are used to analyze the performance of the patterns in the class. The indices of the representatives are shown with boldface in column 4. The fifth column contains vectors $\aa$ defining the PBER for the ABD in~\eqref{eq:BER_maxlog_new}. The patterns are ordered from best to worst PBER for high SNR, as predicted by the vectors $\aa$. All $4!=24$ valid labelings for $4$-PAM give three vectors $\qq$, i.e., three labelings that give different BER: $\qq=[1,2]$ (BRGC), $\qq=[1,3]$ (NBC), and $\qq=[2,3]$ (AGC). These labelings are listed in the first part of \tabref{tab:MPAM_labelings} together with their class indices $\qq$, pattern indices $\setW$, and vectors $\aalpha$ defining the BER for the ABD. The labelings are also ordered from best to worst BER for high SNR as predicted by the vectors $\aalpha$.
\end{example}

\begin{table}[t]
    \renewcommand{\arraystretch}{1.2}
\small
    \centering
        \caption{Classes of patterns for $4$-PAM with their corresponding class indices $q$, representatives $\pp$, types, decimal representations of the patterns $w$, vectors $\aa$ defining the PBER for the ABD, and  thresholds for the representatives}
        \begin{tabular}{@{~}c@{~}|@{~}c@{~}|@{~}c@{~}|@{~}c@{~}|@{~}c@{~}|@{~}l@{~}}
            \hline
            $q$ & $\pp$&  Type & $w$& $\aa$ &  Thresholds \\
            \hline

            \hline
            $1$ & $[0,0,1,1]$&ARE&${\bf3} \tab 12$ & $[2, \tab\tabb2, \tabb\tab0]$& $\beta_2 = 0$ \\
            \hline
            $2$ & $[0,1,1,0]$&RE&${\bf6} \tab\tab 9$ & $[4, \tabb\tab2, -2]$ & $\beta_3 = -\beta_1$\\
            \hline
            $3$ & $[0,1,0,1]$&ARE&${\bf5}\tab 10 $& $[6, -4, \tab\tabb2]$ & $\beta_3 = -\beta_1$, $\beta_2 = 0$ \\
            \hline

            \hline
        \end{tabular}
        \label{tab:4PAM_seq}
\end{table}

The next theorem gives closed form expressions for the number of classes for length-$M$ patterns.
\begin{theorem}\label{the:number_classes}
For $M$-PAM, all the length-$M$ bit patterns can be grouped into $Q$ classes, where
\begin{align}
    Q &= \frac{1}{4}\left(\tbinom{M}{M/2} + \tbinom{M/2}{M/4}+ 2^{M/2}\right),\label{eq:Q}
\end{align}
among which $Q_{\mathrm{RE}}$ classes have only RE patterns, $Q_{\mathrm{ARE}}$ only ARE patterns, and $Q_{\mathrm{ASY}}$ only asymmetric patterns, where
\begin{align}
    Q_{\mathrm{RE}} &= \frac{1}{2}\tbinom{M/2}{M/4},\label{eq:Q_refl}\\
    Q_{\mathrm{ARE}} &= 2^{M/2-1},\label{eq:Q_arefl}\\
    Q_{\mathrm{ASY}} &= \frac{1}{4}\left(\tbinom{M}{M/2} - \tbinom{M/2}{M/4} - 2^{M/2}\right). \label{eq:Q_asym}
\end{align}
\end{theorem}
\begin{IEEEproof}
Any pattern $\pp$ must contain $M/2$ zeros and $M/2$ ones, hence, the total number of patterns is equal to $\tbinom{M}{M/2}$. For a pattern to be RE, $p_i = p_{M-i+1}$, i.e., the positions of the $M/4$ ones in $[p_1,\dots,p_{M/2}]$ fully describe the pattern, and thus, the number of RE patterns is $\tbinom{M/2}{M/4}$. There are two members in every class of RE patterns, $\pp = \refl(\pp)$ and $\inv(\refl(\pp)) = \inv(\pp)$, which gives~\eqref{eq:Q_refl}.

For a pattern to be ARE, $p_i = \bar{p}_{M-i+1}$, i.e., the positions of the ones in $[p_1,\dots,p_{M/2}]$ fully describe the pattern where the number of ones in $[p_1,\dots,p_{M/2}]$ is between 0 and $M/2$. From that, it follows that there are $2^{M/2}$ ARE patterns. There are two members in every class of ARE patterns ($\pp = \inv(\refl(\pp))$ and $\refl(\pp) = \inv(\pp)$), which gives~\eqref{eq:Q_arefl}.

All the remaining classes include only asymmetric patterns. The number of asymmetric patterns can be obtained by subtracting $2Q_{\mathrm{RE}}$ and $2Q_{\mathrm{ARE}}$ from the total number of patterns, i.e., $\tbinom{M}{M/2}-2Q_{\mathrm{RE}}-2Q_{\mathrm{ARE}}$. There are four patterns in each class, as  $\pp \neq \refl(\pp)$ and $\pp \neq  \refl(\inv(\pp))$ (or equivalently, $ \refl(\pp) \neq  \inv(\pp)$). Using this, \eqref{eq:Q_asym} is obtained.

Finally, the total number of classes in \eqref{eq:Q} is obtained as $Q_{\mathrm{RE}}+Q_{\mathrm{ARE}}+Q_{\mathrm{ASY}}$.
\end{IEEEproof}

\begin{table}[t]
    \renewcommand{\arraystretch}{1.2}
    \centering
    \small
    \caption{Some common labelings for $4$-PAM and $8$-PAM with their corresponding class indices $\qq$, patterns indices $\setW$, and vectors $\aalpha$ defining the BER for the ABD}
        \begin{tabular}{@{~}c@{~}|@{~}c@{~}|@{~}c@{~}|@{~}c@{~}|@{~}l@{~}}
            \hline
            $M$ & Labeling & $\qq$ & $\setW$ & $\aalpha$ \\
            \hline

            \hline
            $4$ &BRGC & $[1, 2]$  & $\{3, 6\}$  & $[6, 4, -2]$ \\
            \hline
            $4$ & NBC  & $[1, 3]$  & $\{3, 5\}$  & $[8, -2, 2]$ \\
            \hline
            $4$ & AGC   & $[2, 3]$  & $\{5, 6\}$  & $[10, -2, 0]$ \\
            \hline
            \hline
            $8$ & BRGC & $[1, 2, 6]$  & $\{15, 60, 102\}$  & $[14, 12, -2, 0, 2, 0, -2]$ \\
            \hline
            $8$ & FBC  & $[1, 2, 10]$  & $\{15, 60, 90\}$   & $[18, 0, 4, 10, -8, 2, -2]$ \\
            \hline
            $8$ & NBC  & $[1, 5, 11]$  & $\{15, 51, 85\}$   & $[22, -4, 8, -10, 8, -2, 2]$ \\
            \hline
            $8$ & BSGC & $[2, 6, 9]$ & $\{105, 60, 102\}$ & $[22, 10, -8, 4, -2, -2, 0]$ \\
            \hline
            $8$ & AGC   & $[9, 10, 11]$ & $\{90, 105, 85\}$  & $[36, -18, 6, 4, -4, -2, 2]$ \\
            \hline

            \hline
        \end{tabular}
        \label{tab:MPAM_labelings}
\end{table}

\theref{the:number_classes} gives the exact number of classes $Q$ for $M$-PAM constellations. A loose bound on this number of classes was previously presented in \cite[eq.~(3.14)]{Stierstorfer09_Thesis}. For $4$-PAM \theref{the:number_classes} gives $Q=3$, $Q_{\mathrm{RE}}=1$, $Q_{\mathrm{ARE}}=2$, and $Q_{\mathrm{ASY}}=0$, which is in agreement with~\exref{ex:4PAM} and~\tabref{tab:4PAM_seq}. The following two examples show the number of classes for $8$-PAM and $16$-PAM.

\begin{example}[Patterns for $8$-PAM]\label{ex:8PAM}
For $8$-PAM ($M = 8$), \theref{the:number_classes} states that there are $Q=23$ classes of patterns, $Q_{\mathrm{RE}}=3$ classes of RE patterns and $Q_{\mathrm{ARE}}=8$ classes of ARE patterns, each containing two patterns. These classes are shown in the first 11 rows of \tabref{tab:8PAM_seq} and ordered from the best to worst PBER for high SNR, as predicted by the vectors $\aa$. The  $Q_{\mathrm{ASY}}=12$ classes of asymmetric patterns, each with four members, are listed in the last 12 rows of \tabref{tab:8PAM_seq} and also ordered in a similar way. By enumerating all $8!=40320$ valid labelings for $8$-PAM, we found 460 vectors $\qq$, i.e., for $8$-PAM there exist only 460 labelings that give different BER. The classes associated to five of the most well-known labelings are: $\qq=[1,2,6]$ (BRGC), $\qq=[1, 2, 10]$ (FBC), $\qq=[1, 5, 11]$ (NBC), $\qq=[2, 6, 9]$ (BSGC), and $\qq=[9,10,11]$ (AGC). These labelings are listed from best to worst BER in the second part of~\tabref{tab:MPAM_labelings}.
\end{example}

\begin{table*}[t]
    \renewcommand{\arraystretch}{1.2}
\small
    \centering
        \caption{Classes of patterns for $8$-PAM with their corresponding class indices $q$, representatives $\pp$, types, decimal representations of the patterns $w$, vectors $\aa$ defining the PBER for the ABD, and  thresholds for the representatives}
        \begin{tabular}{@{~}c@{~}|@{~}c@{~}|@{~}c@{~}|@{~}c@{~}|@{~}c@{~}|@{~}l@{~}}
            \hline
            $q$ & $\pp$&  Type & $w$& $\aa$ &  Thresholds \\
            \hline

            \hline
            $1$ & $[0,0,0,0,1,1,1,1]$ &ARE&$\bf{15}$ $240$&$[\tab2,\tab 2,\tab\, 2,\tab 2,\tab 0,\,\tab 0,\tab 0]$& $\beta_4  = 0$\\
            \hline
            $2$ & $[0,0,1,1,1,1,0,0]$ &RE&${\bf60}$ $195$ &$[\tab4,\tab4,\tab2,\tabb2,-2,-2,\tab0]$& $\beta_6 = -\beta_2 = f(t_2)$\\
            \hline
            $3$ & $[1,1,1,0,1,0,0,0]$ &ARE&$23$  $\bf{232}$  &$[\tab6, -2,\tab 2,\tab 0,\tab 2,\tab 0,\tab 0]$  &$\beta_5 = -\beta_3 = f(t_1)$, $\beta_4 = 0$\\
            \hline
            $4$ & $[0,1,1,1,0,0,0,1]$&ARE& ${\bf113}$ $142$ & $[\tab6, \tabb4,\tabb 4, -4, -2, -2,\tabb 2]$ &$\beta_7 = -\beta_1 = f(t_2)$, $\beta_4 = 0$\\
            \hline
            $5$ & $[0,0,1,1,0,0,1,1]$ &ARE&${\bf51}$ $204$ & $[\tab6,\tab6,-4,-4,\tab2,\tabb2,\tab0]$&$\beta_6 = -\beta_2 = f(t_2)$, $\beta_4 = 0$\\
            \hline
            $6$ & $[0,1,1,0,0,1,1,0]$ &RE&${\bf102}$ $153$ & $[\tab8,\tabb 6, -6, -4,\tabb 4,\tabb 2, -2]$ &$\beta_7 = -\beta_1 = f(t_2)$, $\beta_5 = -\beta_3 = f(t_3)$\\
            \hline
            $7$ & $[0,0,1,0,1,0,1,1]$ &ARE&$\bf{43}$ $212$& $[10, -6,\tab 4, -2,\tabb 0,\tab 2,\tab 0]$ &$\beta_6 = -\beta_2 = f(t_2)$, \\
              &              &       &   &&$\beta_5 = -\beta_3 = f(t_3)$,  $\beta_4 = 0$\\
            \hline
            $8$ & $[0,1,0,0,1,1,0,1]$ &ARE&${\bf77}$ $178$ & $[10,\tabb 0, -6,\tab 2,\tab 4, -4,\tab 2]$  &$\beta_7 = -\beta_1 = f(t_2)$,\\
               &               &      &   &  &$\beta_6 = -\beta_2 = f(t_3)$, $\beta_4 = 0$\\
            \hline
            $9$ & $[0,1,1,0,1,0,0,1]$&ARE& ${\bf105}$ $150$& $[10, \tabb 0, -4, \tabb 6, -4, -2, \tabb 2]$&$\beta_7 = - \beta_1 = f(t_2)$,\\
               &                &  & &      &$\beta_5 = - \beta_3 =  f(t_3)$, $\beta_4 = 0$\\
            \hline
            $10$ & $[1,0,1,0,0,1,0,1]$ &RE&$90$ ${\bf165}$ & $[12, -6,\tabb 0,\tabb 6, -6, \tabb 4, -2]$&$\beta_7 = -\beta_1 = f(t_1)$, \\
               &                 &   &    & &$\beta_6 = -\beta_2 = f(t_3)$, $\beta_5 = -\beta_3 = f(t_2)$\\
            \hline
            $11$ & $[0,1,0,1,0,1,0,1]$ &ARE&${\bf85}$ $170$ & $[14, -12, 10, -8, 6, -4, 2]$ &$\beta_7 = -\beta_1 = f(t_2)$, $\beta_6 = -\beta_2 = f(t_3)$,\\
               &                &   &   & &$\beta_5 = -\beta_3 = f(t_1)$, $\beta_4 = 0$\\
            \hline
            \hline
            $12$ & $[0,0,0,1,1,1,1,0]$&ASY& ${\bf30}$ $120$ $135$ $225$& $[\tab4,\tabb 3,\tabb 3,\tabb 2, -2, -1, -1]$ &$\beta_3, \beta_7$\\
            \hline
            $13$ & $[0,0,0,1,1,1,0,1]$&ASY& ${\bf29}$ \hspace{0.5mm} $71$ $184$ $226$& $[\tab6, \tab 1, \tab2, -3, \tab 1,\tab 0,\tab 1]$ &$\beta_3, \beta_6,\beta_7$\\
            \hline
            $14$ & $[0,0,0,1,1,0,1,1]$&ASY& ${\bf27}$\hspace{1.5mm} $39$ $216$ $228$ & $[\tab6, \tab 2, -3, \tab1,\tab 1, \tab1,\tab 0]$ &$\beta_3, \beta_5, \beta_6$ \\
            \hline
            $15$ & $[0,0,1,1,1,0,0,1]$&ASY& ${\bf57}$ \hspace{0.5mm} $99$ $156$ $198$ & $[\tab6, \tab5, \tabb 0, -3, -3, \tab2, \tab1]$ &$\beta_2, \beta_5, \beta_7$\\
            \hline
            $16$ & $[0,0,1,0,1,1,1,0]$&ASY& ${\bf46}$ $116$ $139$ $209$& $[\tab8, -1, 2, -1,\, 3, -2, -1]$ &$\beta_2, \beta_3, \beta_4, \beta_7$\\
            \hline
            $17$ & $[0,0,1,1,1,0,1,0]$&ASY& ${\bf58}$ \hspace{0.5mm} $92$ $163$ $197$ & $[\tab8, -1, 3, -2,\, 2, -1, -1]$ &$\beta_2, \beta_5, \beta_6, \beta_7$\\
            \hline
            $18$ & $[0,1,0,0,1,1,1,0]$&ASY& ${\bf78}$ $114$ $141$ $177$& $[\tab8, \,2, -1, -1, -1, 3, -2]$&$\beta_1, \beta_2, \beta_4, \beta_7$\\
            \hline
            $19$ & $[0,0,1,1,0,1,1,0]$&ASY& ${\bf54}$ $108$ $147$ $201$ & $[\tab8, \tabb3, -6,\tabb 3,\tabb 3, -2, -1]$ &$\beta_2, \beta_4, \beta_5, \beta_7$\\
            \hline
            $20$ & $[0,0,1,0,1,1,0,1]$&ASY& ${\bf45}$ \hspace{0.5mm} $75$ $180$ $210$& $[10, -3, -3, \tabb6, -4,\tabb 1,\tabb 1]$ &$\beta_2, \beta_3, \beta_4, \beta_6, \beta_7$\\
            \hline
            $21$ & $[0,0,1,1,0,1,0,1]$&ASY& ${\bf53}$ \hspace{0.5mm} $83$ $172$ $202$  & $[10, -3, \tab1, \tab0, -2, \tab1,\tabb 1]$ &$\beta_2, \beta_4, \beta_5, \beta_6, \beta_7$\\
            \hline
            $22$ & $[0,1,0,1,1,0,0,1]$&ASY& ${\bf89}$ $101$ $154$ $166$ &$[10, \tab0, -3, \tab1,\tab 1, -3,\tabb 2]$&$\beta_1, \beta_2, \beta_3, \beta_5, \beta_7$\\
            \hline
            $23$ & $[0,1,0,1,0,1,1,0]$&ASY& ${\bf86}$ $106$ $149$ $169$& $[12, -6, 3, -1, -1, \,3, -2]$&$\beta_1, \beta_2, \beta_3, \beta_4, \beta_5, \beta_7$\\
            \hline

            \hline
        \end{tabular}
        \label{tab:8PAM_seq}
\end{table*}

\begin{example}[Patterns for $16$-PAM]\label{ex:16PAM}
For $16$-PAM ($M = 16$) \theref{the:number_classes} states that there are $Q  = 3299$ classes, $Q_{\mathrm{RE}}=35$ of the classes contain RE patterns, $Q_{\mathrm{ARE}}=128$ classes contain ARE patterns, and $Q_{\mathrm{ASY}}=3136$ classes contain ASY patterns. The patterns for $16$-PAM are not listed in this paper.
\end{example}


\section{Thresholds for the BD}\label{sec:Thresholds}

In this section, we show that the thresholds for the BD can be found by solving an $(M-1)$th power polynomial equation and give a closed-form solutions for $4$-PAM and $8$-PAM with RE and ARE patterns.

\subsection{Threshold Computation}\label{sec:opt_general}

The problem of finding the  thresholds $\beta_k$ for the BD is equivalent to finding the solutions of $l(y)=0$. In the following theorem we show how this can be done for $M$-PAM constellations.

\begin{theorem}\label{the:how_to_find_thresholds}
    The  thresholds for the BD and $M$-PAM constellations with a pattern $\pp$ are
    \begin{equation}
        \beta_k = \frac{1}{4\SNR d}\log{z_n}, \label{eq:threshold}
    \end{equation}
    where $z_n$ are the real and positive solutions of
    \begin{equation}
    \sum_{i = 1}^{M}{\check{p}_{i}A^{\frac{(M/2 - i)(M/2+1-i)}{2}} z^{i-1}} = 0,
    \label{eq:expression_general}
    \end{equation}
    and
    \begin{equation}
        A = \mathrm{e}^{-8\SNR d^2} \label{eq:substitution}.
    \end{equation}
\end{theorem}
\begin{IEEEproof}
Using~\eqref{eq:L-value}, $l(y)=0$ is equivalent to $\mathrm{e}^{l(y)}=1$, which can be restated as
\begin{equation}
    h(y) \triangleq \sum_{i = 1}^{M}{\check{p}_i{\mathrm{e}^{-\SNR(y + d(M - 2i + 1))^2}}}, \label{eq:ber_general_deriv1}
\end{equation}
where the definition of of the $M$-PAM symbols was used. Factorizing~\eqref{eq:ber_general_deriv1} gives
\begin{multline}
    h(y) = \mathrm{e}^{-\SNR y^2 - 2\SNR d y (M-1) - \SNR d^2} \\
    \cdot\sum_{i = 1}^{M}{\check{p}_i\mathrm{e}^{4\SNR d y (i-1)} \mathrm{e}^{-8 \SNR d^2 \frac{(M/2 - i)(M/2+1-i)}{2}}}. \label{eq:noname2}
\end{multline}


Using~\eqref{eq:substitution} in \eqref{eq:noname2} together with substitution $z = \mathrm{e}^{4\SNR d y}$, \eqref{eq:expression_general} is obtained by setting $h(y)=0$ and removing the nonzero factor preceding the summation in~\eqref{eq:noname2}. The expression in \eqref{eq:expression_general} is an $(M-1)$-power polynomial\footnote{Interestingly, for $i=1,\ldots,M-1$ with $i\neq M/2$ and $i\neq M/2+1$, the powers of $A$ in \eqref{eq:expression_general} are the so-called ``triangular numbers''. }, and thus, it always has $M-1$ solutions. Because of the substitution $z = \mathrm{e}^{4\SNR d y}$, only the positive (and real) roots need to be considered.
\end{IEEEproof}

\begin{figure}[t]
\newcommand{\scale}{0.9}
\newcommand{\scalesmall}{0.7}
\begin{center}
    \psfrag{SNR}[tc][tc]{$\SNR$~[dB]}
    \psfrag{BER}[bc][Bc][\scale]{}
    \psfrag{ s1}[cl][cl][\scalesmall]{$s_1$}
    \psfrag{ s2}[cl][cl][\scalesmall]{$s_2$}
    \psfrag{ s3}[cl][cl][\scalesmall]{$s_3$}
    \psfrag{ s4}[cl][cl][\scalesmall]{$s_4$}
    \psfrag{ s5}[cl][cl][\scalesmall]{$s_5$}
    \psfrag{ s6}[cl][cl][\scalesmall]{$s_6$}
    \psfrag{ s7}[cl][cl][\scalesmall]{$s_7$}
    \psfrag{ s8}[cl][cl][\scalesmall]{$s_8$}
    \psfrag{ s9}[cl][cl][\scalesmall]{$s_9$}
    \psfrag{ s10}[cl][cl][\scalesmall]{$s_{10}$}
    \psfrag{ s11}[cl][cl][\scalesmall]{$s_{11}$}
    \psfrag{ s12}[cl][cl][\scalesmall]{$s_{12}$}
    \psfrag{ s13}[cl][cl][\scalesmall]{$s_{13}$}
    \psfrag{ s14}[cl][cl][\scalesmall]{$s_{14}$}
    \psfrag{ s15}[cl][cl][\scalesmall]{$s_{15}$}
    \psfrag{ s16}[cl][cl][\scalesmall]{$s_{16}$}
    \includegraphics[width=0.95\columnwidth]{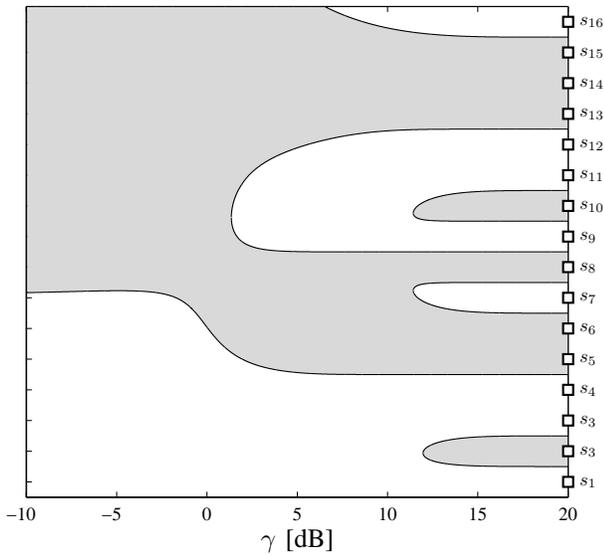}
    \caption{Thresholds $\beta_k$ in~\eqref{eq:threshold} for $16$-PAM with the ASY pattern $\pp_{45745}= [1,0,1,1,0,0,1,0,1,0,1,1,0,0,0,1]$ obtained by solving~\eqref{eq:expression_general} numerically. The constellation points are shown with squares. Gray and white areas indicate $\mathcal{Y}_{0}$ and $\mathcal{Y}_{1}$, resp.}
    \label{fig:thresh16pam}
\end{center}
\end{figure}

\theref{the:how_to_find_thresholds} gives a general expression for finding the thresholds for $M$-PAM with any pattern $\pp$. After finding the roots of \eqref{eq:expression_general}, the  thresholds $\beta_k$ may easily be obtained from \eqref{eq:threshold}. The main problem is that finding the roots of \eqref{eq:expression_general} does not admit simple closed-form solutions. However, it can always be solved numerically. \figref{fig:thresh16pam} illustrates the  thresholds for $16$-PAM with the pattern $\pp_{45745} = [1,0,1,1,0,0,1,0,1,0,1,1,0,0,0,1]$ obtained by solving~\eqref{eq:expression_general} numerically. In the following two sections we show how this problem can be solved analytically for $4$-PAM and $8$-PAM with RE or ARE patterns.

\subsection{Thresholds for $4$-PAM}\label{sub:opt_dem_4PAM}
The following theorem shows how the  thresholds are found for $4$-PAM with any pattern.
\begin{theorem}
\label{theor:4pam_thresh}
The  thresholds $\beta_k$ for any pattern for $4$-PAM, as listed in the last column in \tabref{tab:4PAM_seq}, can be expressed as
\begin{align}
    \beta_1 & = -\beta_3,\label{eq:4pam_betas.1}\\
    \beta_2 & = 0,\label{eq:4pam_betas.2}\\
    \beta_3 & = \frac{1}{4\SNR d}\log{\left|\frac{1 \!+\!  \check{p}_1 \check{p}_4 A\! +\! \sqrt{(1\! +\! \check{p}_1 \check{p}_4 A)^2 - 4A^2}}{2A}\right|} ,\label{eq:4pam_betas.3}
\end{align}
 where $A$ is given by~\eqref{eq:substitution}.
\end{theorem}
\begin{IEEEproof}
The proof is given in Appendix~\ref{Appendix.theor:4pam_thresh}.
\end{IEEEproof}

\theref{theor:4pam_thresh} gives closed-form expressions for any pattern for $4$-PAM, and thus, it allows us to compute the BER for all the labelings in the first part of \tabref{tab:MPAM_labelings}. The results in \theref{theor:4pam_thresh} can be shown to coincide to those in \cite[eq.~(10)]{Simon05feb} when the BRGC is considered.


\subsection{Thresholds for $8$-PAM}\label{sub:opt_dem_8PAM}

The following theorem shows how the  thresholds may be found for $8$-PAM with RE or ARE patterns. These patterns are of great value because all the most commonly studied labelings (e.g., BRGC, NBC, FBC, BSGC, AGC) can be composed from them (cf.~\tabref{tab:MPAM_labelings}).

\begin{theorem}\label{theor:8pam_thresh}
The  thresholds $\beta_k$ for the patterns in the classes $q = 1,2,\dots,11$ for $8$-PAM can be expressed as
\begin{equation}
    \beta_k =  -\beta_{8 - k} =
\begin{cases}
    f(t_n)& \text{ if } k = 5,6,7,\\
    0& \text{ if }  k = 4,
\end{cases}
    \label{eq:threshold2}
\end{equation}
\begin{equation}
    f(t) \triangleq \frac{1}{4\SNR d}\log{\left|\frac{|t| + \sqrt{|t|^2-4}}{2}\right|},
\label{eq:function}
\end{equation}
with
\begin{align}
\label{eq:t1.1}
    	& t_1 \triangleq \frac{1}{6\check{p}_1A^3} \biggl(T + \frac{2\sqrt[3]{2}C}{B} + \sqrt[3]{4}B \biggr),\\
\label{eq:t1.2}
    &t_2 \triangleq \frac{1}{6\check{p}_1A^3} \biggl(T - \frac{\sqrt[3]{2}(1+\sqrt{3}\jmath)C}{B} - \frac{1-\sqrt{3}\jmath}{\sqrt[3]{2}}B\biggr),\\
\label{eq:t1.3}
    &t_3 \triangleq \frac{1}{6\check{p}_1A^3} \biggl(T - \frac{\sqrt[3]{2}(1-\sqrt{3}\jmath)C}{B} - \frac{1+\sqrt{3}\jmath}{\sqrt[3]{2}}B \biggr),
\end{align}
$A$ is given by~\eqref{eq:substitution},
\begin{align*}
	T \triangleq &2(\check{p}_8A^3-\check{p}_2),\\
    B \triangleq &\sqrt[3]{\sqrt{{D}^2 - 4{C}^3} -  \check{p}_1 \check{p}_8 D},\\
    C \triangleq  &7A^6 + \check{p}_2\check{p}_8A^3-3\check{p}_1\check{p}_3A + 1,\\
    D \triangleq &7\check{p}_1A^9 - 12\check{p}_1 \check{p}_2\check{p}_8A^6 - 18\check{p}_3A^4 \\
    	&\qquad + 3\check{p}_1(1 + 9\check{p}_4\check{p}_8)A^3 - 9\check{p}_2\check{p}_3\check{p}_8A + 2\check{p}_1 \check{p}_2\check{p}_8,
\end{align*}
and the relationship between $k$ and $n$ in \eqref{eq:threshold2} for the different classes $q$ is listed in the last column of \tabref{tab:8PAM_seq}. As the relationship between $n$ and $k$ depends on the particular pattern, the representative of the class (the second column of \tabref{tab:8PAM_seq}) should be used in the presented equations.
\end{theorem}
\begin{IEEEproof}
The proof is given in Appendix~\ref{Appendix.theor:8pam_thresh}.
\end{IEEEproof}

\theref{theor:8pam_thresh} shows how to analytically obtain the  thresholds for the BD with $8$-PAM and any RE or ARE pattern, for instance, thresholds shown in~\figref{fig:lvalues8PAM}. Using these results,  the PBER can be calculated using~\eqref{eq:BER_general}, which gives PBER expressions for $11$ out of $23$ classes, or equivalently, for $56$ different labelings, including the 5 shown in the second part of \tabref{tab:MPAM_labelings}.

\begin{remark}\label{4PAM.Asymptotic}
In the high SNR regime, i.e., $\SNR\to\infty$, all the thresholds in~\thesref{theor:4pam_thresh}{theor:8pam_thresh} tend to midpoints, i.e., the same constant thresholds used in the ABD for all SNR. This fact can easily be proven analytically for $4$-PAM by evaluating $\lim_{\SNR \to \infty}\beta_3$ and applying l'H\^opital's rule. For $8$-PAM a similar proof exists, however, in this case it is not straightforward due to the complexity of the threshold expressions. These results can be intuitively understood from the fact that the max-log approximation in~\eqref{eq:max_log_rrl} becomes more precise when the SNR increases, and hence, the thresholds for the BD and ABD are expected to coincide when $\SNR\to\infty$.
\end{remark}

\section{Numerical Results}\label{sec:numerical_results}

\begin{figure}[t]
\newcommand{\scale}{0.9}
\begin{center}
    \psfrag{SNR}[tc][tc]{$\SNR$~[dB]}
    \psfrag{BER}[bc][Bc][\scale]{}
    \psfrag{d211111}[cl][cl][\scale]{$\beta_7 = -\beta_1$}
    \psfrag{d22}[cl][cl][\scale]{$\beta_6=-\beta_2$}
    \psfrag{d33}[cl][cl][\scale]{$\beta_5 = -\beta_3$}
    \psfrag{ s1}[cl][cl][\scale]{$s_1$}
    \psfrag{ s2}[cl][cl][\scale]{$s_2$}
    \psfrag{ s3}[cl][cl][\scale]{$s_3$}
    \psfrag{ s4}[cl][cl][\scale]{$s_4$}
    \psfrag{ s5}[cl][cl][\scale]{$s_5$}
    \psfrag{ s6}[cl][cl][\scale]{$s_6$}
    \psfrag{ s7}[cl][cl][\scale]{$s_7$}
    \psfrag{ s8}[cl][cl][\scale]{$s_8$}
    \psfrag{ b1}[cl][cl][\scale]{$\betat_1$}
    \psfrag{ b2}[cl][cl][\scale]{$\betat_2$}
    \psfrag{ b3}[cl][cl][\scale]{$\betat_3$}
    \psfrag{ b4}[cl][cl][\scale]{}
    \psfrag{ b5}[cl][cl][\scale]{$\betat_5$}
    \psfrag{ b6}[cl][cl][\scale]{$\betat_6$}
    \psfrag{ b7}[cl][cl][\scale]{$\betat_7$}
    \includegraphics[width=0.95\columnwidth]{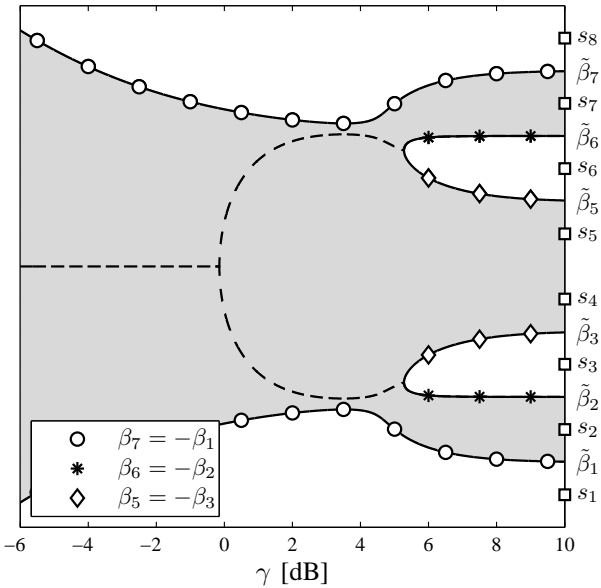}
    \caption{The thresholds  for $\pp_{165}=[1,0,1,0,0,1,0,1]$ ($q = 10$) for $8$-PAM in \theref{theor:8pam_thresh} vs. SNR. Virtual thresholds are shown with dashed lines. The  thresholds for the ABD $\betat_k$ and the constellation points (squares) are also shown. Gray and white areas indicate $\mathcal{Y}_{0}$ and $\mathcal{Y}_{1}$, resp.}
    \label{fig:thresh_all}
\end{center}
\end{figure}

In~\figref{fig:thresh_all} we show the thresholds given by \theref{theor:8pam_thresh} for the pattern $\pp_{165}$ ($q = 10$) for $8$-PAM. The figure is symmetric with respect to zero due to the symmetry of the pattern. At $\SNR\approx5.3$~dB the pairs of thresholds $\beta_2$ and $\beta_3$, and $\beta_5$ and $\beta_6$ merge and become virtual for all $\SNR<5.3$ dB. All the virtual thresholds shown with dashed lines satisfy the conditions in \theref{theo:ber_1D_general.virtual}. As expected (see Remark~\ref{4PAM.Asymptotic}), when $\SNR\rightarrow\infty$, the BD thresholds coincide with the ABD thresholds.

\begin{figure}[t]
\newcommand{\scale}{0.9}
\begin{center}
    \psfrag{SNR}[tc][tc]{$\SNR$~[dB]}
    \psfrag{BER}[bc][Bc][\scale]{PBER}
    \psfrag{d1111}[cl][cl][\scale]{$q = 3$}
    \psfrag{d2}[cl][cl][\scale]{$q = 10$}
    \psfrag{d3}[cl][cl][\scale]{$q = 16$}
    \psfrag{d4}[cl][cl][\scale]{$q = 22$}
%

    \includegraphics[width=0.95\columnwidth]{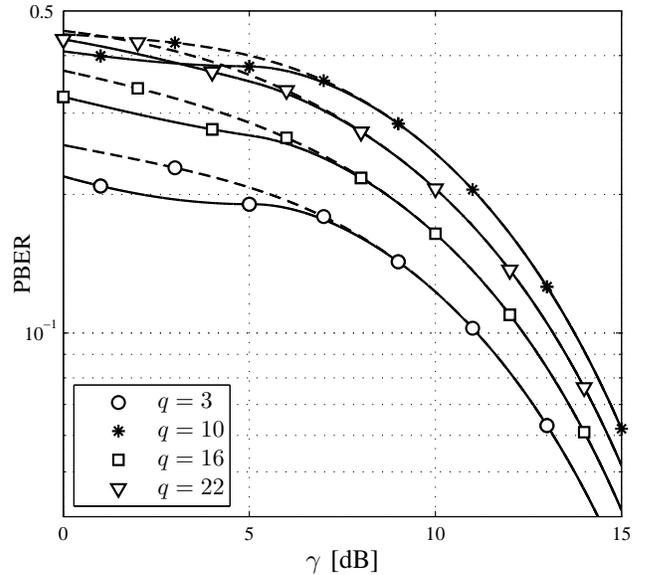}
    \caption{The PBER for $8$-PAM with ARE ($q = 3$) and RE ($q=10$) patterns and ASY patterns ($q = 16, 22$). Solid lines correspond to the BD and dashed lines to the ABD. The threshold for the BD were obtained using \theref{theor:8pam_thresh} for $q = 3, 10$ (the thresholds for $q = 10$ are shown in~\figref{fig:thresh_all}) and solving~\eqref{eq:expression_general} numerically for~$q = 16, 22$.}
    \label{fig:pam8all_seq}
\end{center}
\end{figure}

\begin{figure}[t]
\newcommand{\scale}{0.9}
\begin{center}
    \psfrag{SNR}[tc][tc]{$\SNR$~[dB]}
    \psfrag{BER}[bc][Bc][\scale]{BER}
    \psfrag{dat5}[cl][cl][\scale]{BRGC}
    \psfrag{dat4}[cl][cl][\scale]{FBC}
    \psfrag{dat3}[cl][cl][\scale]{NBC}
    \psfrag{dat2}[cl][cl][\scale]{BSGC}
    \psfrag{dat111}[cl][cl][\scale]{AGC}
    \includegraphics[width=0.95\columnwidth]{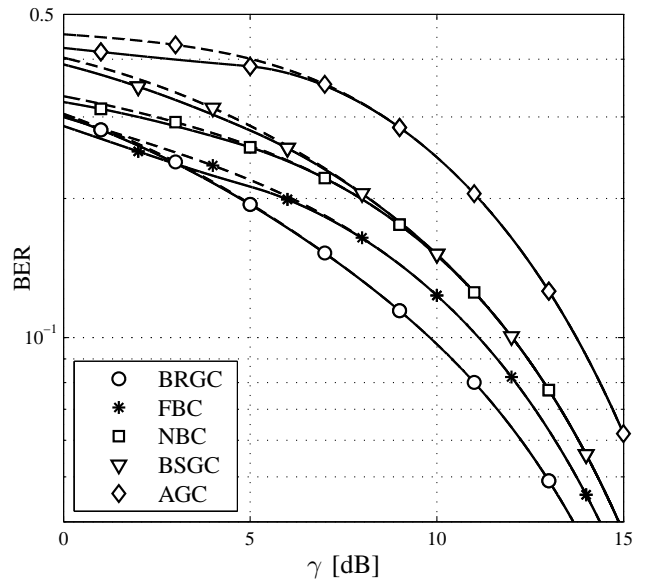}
    \caption{The BER for $8$-PAM with the 5 labelings in~\tabref{tab:MPAM_labelings}. Solid lines correspond to the BD and dashed lines to the ABD.}
    \label{fig:pam8_labelings}
\end{center}
\end{figure}

The PBER for $8$-PAM with some selected patterns from~\tabref{tab:8PAM_seq} using \eqref{eq:BER_general} is presented in~\figref{fig:pam8all_seq}. The thresholds are calculated analytically for $q = 3, 10$ and numerically for $q = 16, 22$. For very low SNR the gap between the BD and the ABD can reach up to several dB, however, this gap decreases when the SNR increases. The same conclusion can be drawn for all other patterns except for $q = 1$, as in this case only one threshold exists $\beta_{M/2}=0$ for all SNR. Hence, for $q = 1$ the BD and the ABD have the same performance for $M$-PAM. To conclude, we present in~\figref{fig:pam8_labelings} the BER for $8$-PAM with the labelings in~\tabref{tab:MPAM_labelings}. From the presented results we conclude that the BD outperforms the ABD, however, for any BER of practical interest (below $0.1$), the difference between the BD and the ABD is negligible.

\section{Conclusions} \label{sec:conclusions}

We proposed a general approach for estimating the performance of the optimal bit-wise demodulator and presented closed-form expressions for the BER for $4$-PAM and $8$-PAM with different labelings. We conclude that a suboptimal symbol-wise demodulator shows no loss compared to the optimal demodulator for all the SNR of interest, which justifies its use in practical systems.

The derived BER expressions for the optimal demodulator can be used to calculate the mutual information (MI) of BICM when the demodulator makes hard decisions on the bits. The optimal bit-wise demodulator does not necessarily maximize the MI. Finding the hard-decision demodulator that maximizes the MI is left for future work.

The proposed technique for finding the zero crossings of the L-values for $8$-PAM works only for reflected or anti-reflected patterns, which includes 11 out of 23 classes of patterns. Extending these results to the remaining classes of patterns for $8$-PAM is left for further investigation as well as generalizing the results to arbitrary $M$.

\appendices

\section{Proof of \theref{theo:ber_1D_general}}\label{Appendix.theo:ber_1D_general}

Let $v_{i,k}$ be the conditional probabilities
\begin{align}
\nonumber
v_{i,1} 	&\triangleq \Pr\{Y \le \beta_1|X=s_i\} \\
		& = 1-\qfunc\left((\beta_1-s_i)\sqrt{2\SNR}\right),\label{eq:v_ik.1}\\
\nonumber
v_{i,k} 	&\triangleq \Pr\{\beta_{k-1} < Y \le \beta_k|X=s_i\} \\
		&= \qfunc\left((\beta_{k-1}-s_i)\sqrt{2\SNR}\right)-\qfunc\left((\beta_k-s_i)\sqrt{2\SNR}\right),\label{eq:v_ik.2}\\
\nonumber
v_{i,M} 	&\triangleq \Pr\{\beta_{M-1} < Y|X=s_i\} \\
		&= \qfunc\left((\beta_{M-1}-s_i)\sqrt{2\SNR}\right), \label{eq:v_ik.3}
\end{align}
where $\beta_k$ for $k \in \setK$ are the thresholds and none of them is virtual. The PBER in~\eqref{eq:P_j} can now be rewritten as
\begin{align}
P &= \frac{1}{M}\sum_{i = 1}^{M}\sum_{k = 1}^{M}e_{i,k}v_{i,k},
\label{eq:P_ev}
\end{align}
where $e_{i,k} \triangleq p_i \oplus p_k\in\{0,1\}$.

Using \eqref{eq:v_ik.1}--\eqref{eq:v_ik.3} the PBER in \eqref{eq:P_ev} can be expressed as
\begin{align}
   P & = \frac{1}{M}\biggl[\sum_{i = 1}^{M}e_{i,1} + \sum_{i = 1}^{M}\sum_{k = 2}^{M}e_{i,k}\qfunc\left((\beta_{k-1}-s_i)\sqrt{2\SNR}\right) \nonumber\\
   	& \qquad\qquad\qquad - \sum_{i = 1}^{M}\sum_{k = 1}^{M-1}e_{i,k}\qfunc\left((\beta_k-s_i)\sqrt{2\SNR}\right)\biggr]\nonumber\\
   & = \frac{1}{2} + \frac{1}{M}\sum_{i = 1}^{M}\sum_{k = 1}^{M-1}(e_{i,k+1}-e_{i,k})\qfunc\left((\beta_k-s_i)\sqrt{2\SNR}\right),
    \label{eq:P_eQ}
\end{align}
where $\sum_{i = 1}^{M}e_{i,1} = \sum_{i = 1}^{M}p_i \oplus p_1 = M/2$ was used.
To obtain the expression in~\eqref{eq:BER_general}, we express $e_{i,k+1}-e_{i,k}$ in \eqref{eq:P_eQ} as
\begin{align}
e_{i,k+1}-e_{i,k} 	&= p_{k+1} \oplus p_{i} - p_{k} \oplus p_{i}\label{eq:g_ik2.1}\\
				& = (p_{k+1}-p_k)(1-2p_i),\label{eq:g_ik2.5}
\end{align}
where the identity $p_i \oplus p_j = p_i\bar{p}_j + \bar{p}_i p_j$ was used together with $\bar{p}_i = 1-p_i$.

\section{Proof of \theref{theor:4pam_thresh}}\label{Appendix.theor:4pam_thresh}

Define the function $h(z)$ as
\begin{equation}
    h(z) = \check{p}_4Az^3 + \check{p}_3z^2 + \check{p}_2z + \check{p}_1A.    \label{eq:4pam_general}
\end{equation}
According to~\theref{the:how_to_find_thresholds}, for $4$-PAM with a pattern $\pp = [p_1, p_2,p_3,p_4]$, equation $h(z) = 0$ needs to be solved in order to find the thresholds. The patterns for $4$-PAM are either RE ($\check{p}_i = \check{p}_{M+1-i}$, $\forall i$) or ARE ($\check{p}_i = - \check{p}_{M+1-i}$, $\forall i$). Therefore
\begin{equation}
    h(z) = \check{p}_1Az^3 + \check{p}_2z^2 \pm \check{p}_2z \pm \check{p}_1A, \label{eq:4pam_symm.relf}
\end{equation}
where the upper and the lower signs correspond to RE and ARE patterns, resp. Using $\check{p}_i^2=1$ and the fact that $\check{p}_1\check{p}_4=\pm1$ for RE and ARE patterns, resp., $h(z)$ can be factorized as
\begin{align}
h(z) = \check{p}_1 (z +\check{p}_1\check{p}_4)(A z^2 + (\check{p}_1\check{p}_2 -A\check{p}_1\check{p}_4)z + A).\label{eq:4pam_fact.relf}
\end{align}

Solving $h(z)=0$ gives the three roots $z_1=-\check{p}_1\check{p}_4$ and
\begin{equation}
        z_{2,3} = \frac{\check{p}_1\check{p}_4 A-\check{p}_1\check{p}_2\pm \sqrt{(\check{p}_1\check{p}_4 A-\check{p}_1\check{p}_2)^2 - 4A^2}}{2A}.\label{eq:4pam_roots.1.arefl}\\
\end{equation}

For $q = 1$ (where $\check{p}_1\check{p}_4=-1$ and $\check{p}_1\check{p}_2=1$) the root $z_1 = 1$, that used in~\eqref{eq:threshold} gives the threshold $\beta_2 = 0$. The other two roots in \eqref{eq:4pam_roots.1.arefl} are complex for all SNR and do not result in thresholds.

In a similar way, for $q = 3$ (where $\check{p}_1\check{p}_2=-1$ and $\check{p}_1\check{p}_4=-1$) $\beta_2 = 0$. When $A \le 1/3$, or equivalently, when $\SNR \ge{5\log{3}}/{8} \approx -1.63 \text{ dB}$, the roots in~\eqref{eq:4pam_roots.1.arefl} are real and positive resulting in thresholds $\beta_1$ and $\beta_3$ by using \eqref{eq:threshold}. When $A > 1/3$ (low SNR), the roots in~\eqref{eq:4pam_roots.1.arefl} are complex and can no longer be used in~\eqref{eq:threshold} for calculating the thresholds. To overcome this, $|z_2|$ and $|z_3|$ are used in the calculation of the thresholds, which together with $\check{p}_1\check{p}_2=-1$ gives~\eqref{eq:4pam_betas.3}. The use of $|\cdot|$ does not affect the result when the roots are real. When the roots are complex, their absolute values are equal to one (can be seen from~\eqref{eq:4pam_roots.1.arefl}), which gives two virtual thresholds $\beta_1 = \beta_3 = 0$ merging with the zero-threshold $\beta_2$ at around $-1.63$ dB. \theref{theo:ber_1D_general.virtual} allows the use of these thresholds in the calculation of the PBER.

Finally, for $q=2$ (where $\check{p}_1\check{p}_2=-1$ and $\check{p}_1\check{p}_4=1$) $z_1=-1$, which results in no threshold. The two roots in \eqref{eq:4pam_roots.1.arefl} are positive for all SNR, resulting in the thresholds $\beta_1$ and $\beta_3$ by using \eqref{eq:threshold}. As the roots are positive the use of $|\cdot|$ does not affect the result, which gives~\eqref{eq:4pam_betas.3}. This completes the proof.

\section{Proof of \theref{theor:8pam_thresh}}\label{Appendix.theor:8pam_thresh}

Define the function $h(z)$ as
\begin{multline}
      h(z) = \check{p}_8A^6z^7 + \check{p}_7A^3z^6 + \check{p}_6Az^5 + \check{p}_5z^4 + \check{p}_4z^3 \\
      + \check{p}_3Az^2 + \check{p}_2A^3z + \check{p}_1A^6.
      \label{eq:PAM8_general}
\end{multline}
According to~\theref{the:how_to_find_thresholds} for $8$-PAM with pattern $\pp = [p_1,p_2,\dots,p_8]$ equation $h(z) = 0$ needs to be solved in order to find the  thresholds. For RE and ARE patterns, $\check{p}_i = \pm \check{p}_{M+1-i}, \tab \forall i$, where the upper and the lower signs correspond to RE and ARE patterns, resp. Using this property, $h(z)$ for RE and ARE patterns is
\begin{multline}
      h(z) = \check{p}_1A^6z^7 + \check{p}_2A^3z^6 + \check{p}_3Az^5 + \check{p}_4z^4 \\
      \pm \check{p}_4z^3 \pm \check{p}_3Az^2 \pm \check{p}_2A^3z \pm \check{p}_1A^6.
      \label{eq:PAM8_symmetric}
\end{multline}
Factorizing~\eqref{eq:PAM8_symmetric} gives
\begin{align}
      &h(z)=(z\pm1)\left(\check{p}_1 A^6 z^6 + \left[\check{p}_2 A^3 \mp \check{p}_1 A^6\right]z^5 \right.\nonumber\\
      &+ \left[\check{p}_1 A^6 \mp \check{p}_2 A^3 + \check{p}_3A \right]z^4 +\left[\mp \check{p}_1 A^6\! + \check{p}_2 A^3\! \mp \check{p}_3 A\! + \check{p}_4 \right]z^3\! \nonumber\\
      &+\left.\! \left[\check{p}_1 A^6 \mp \check{p}_2 A^3\! +\! \check{p}_3 A \right]z^2 +\left[\check{p}_2 A^3 \mp \check{p}_1 A^6 \right]z + \check{p}_1 A^6\right).
      \label{eq:lvalto7}
\end{align}
Rearranging the terms in~\eqref{eq:lvalto7} $h(z)$ can be written as
 \begin{align}
      &h(z) = z^3(z\pm1)\left(\check{p}_1 A^6 (z^3 + z^{-3}) + \left[\check{p}_2 A^3 \mp \check{p}_1 A^6\right](z^2 + z^{-2})\right. \nonumber\\
      &+\left[\check{p}_1 A^6 \mp \check{p}_2 A^3 + \check{p}_3A \right](z^1 +z^{-1}) \nonumber\\
      &\left.+\left[\mp \check{p}_1 A^6 + \check{p}_2 A^3 \mp \check{p}_3 A + \check{p}_4 \right]\right).
      \label{eq:lvalto71}
\end{align}
Using the substitution
\begin{equation}
    t(z) = z + z^{-1},
    \label{eq:substitution2}
\end{equation}
\eqref{eq:lvalto71} can be modified to
\begin{align}
    &h(z) = z^3(z\pm1)\left(\check{p}_1A^6t(z)^3 + \left[\check{p}_2A^3 \mp \check{p}_1A^6\right]t(z)^2\right. \nonumber\\
     &+\left[-2\check{p}_1A^6 \mp \check{p}_2A^3 + \check{p}_3A\right]t(z)\nonumber\\
     &\left.+\left[\pm\check{p}_1A^6-\check{p}_2A^3 \mp \check{p}_3A + \check{p}_4\right]\right).
    \label{eq:lvalto8}
\end{align}

Finding positive roots of $h(z)=0$ can now be done analytically. For ARE patterns the second factor in~\eqref{eq:lvalto8} gives a root equal to one resulting in $\beta_4 = 0$.
The roots of the last factor in~\eqref{eq:lvalto8} need to be found. As a first step we solve it with respect to $t(z)$, where the roots $t_n$ are shown in~\eqref{eq:t1.1}--\eqref{eq:t1.3}, where $\pm1$ was replaced by $+\check{p}_1 \check{p}_8$ to distinguish between RE and ARE patterns. As $z$ should be real and positive, only real and positive $t_n$ need to be considered. Two out of three roots $t_n$ may combine into a complex conjugated couple, but the third root is always real. Every positive root $t_n$ gives two roots for $z$ in \eqref{eq:lvalto7}, which can be found from~\eqref{eq:substitution2} as
\begin{equation}
\label{eq:noname4}
    z_{2n-1,2n}=\frac{t_n\pm \sqrt{{t_n}^2-4}}{2},
\end{equation}
where $z_{2n-1}=1/z_{2n}$. Due to~\eqref{eq:threshold} and the symmetry of the patterns, these two roots give the two thresholds $\beta_k = -\beta_{8-k}$, which justifies the first equality in~\eqref{eq:threshold2}. When $t_n$ is real and $t_n \ge 2$, the roots in \eqref{eq:noname4} are positive and give thresholds $\beta_k= -\beta_{8-k}$. Because of $t_n$ is real and the roots $z_{2n-1}$ and $z_{2n}$ are positive, the use of $|\cdot|$ (three times) in \eqref{eq:function} does not change the result. By analyzing all the roots $t_n$, the thresholds were found and listed in \tabref{tab:8PAM_seq}, where the last column shows the relation between the  threshold $\beta_k$ and the roots $t_n$ shown in \eqref{eq:t1.1}--\eqref{eq:t1.3}.

For the listed thresholds in~\tabref{tab:8PAM_seq}, $t_n$ in~\eqref{eq:t1.1}--\eqref{eq:t1.3} is never real and negative, however $t_n$ can be either complex or real with $0\le t_n<2$ for some $\SNR < \SNR_0$, resulting in virtual thresholds $\beta_k$ and $\beta_{8-k}$. In what follows, we show that these thresholds are equal to each other when using \eqref{eq:function}, i.e., they fulfill the conditions in \theref{theo:ber_1D_general.virtual}. First, consider the case when $t_n$ is real but $0\le t_n<2$. In this case $z_{2n-1}$ and $z_{2n}$ are complex with unit magnitude and according to~\eqref{eq:threshold2}, the corresponding thresholds $\beta_k = -\beta_{8-k}$ are equal to zero. By analyzing the thresholds for all the RE and ARE patterns, we find that $\beta_5 = -\beta_3$ for $q = 3,6,9,11$ and $\beta_6 = -\beta_2$ for $q = 5$, which are separated by either no threshold or by the threshold $\beta_4 = 0$. These thresholds can be used in~\eqref{eq:BER_general} according to \theref{theo:ber_1D_general.virtual}.

Second, consider the case when $t_n$ is complex. One of the other two roots of~\eqref{eq:lvalto8} $t_{n'}$ giving $\beta_{k'} = -\beta_{8- k'}$ is such that $t_{n'} = {t_n}^*$, which means that $|t_n| = |t_{n'}|$. This leads to two pairs of the thresholds $\beta_k = \beta_{k'}$ and $\beta_{8-k} = \beta_{8-k'}$. Revising the thresholds for all the RE and ARE patterns we conclude that corresponding thresholds are: $\beta_6 = -\beta_2$ and $\beta_7 = -\beta_1$ for $q = 8,11$ and $\beta_5 = -\beta_3$ and $\beta_6 = -\beta_2$ for $q = 7,10$. These thresholds can be used in~\eqref{eq:BER_general} according to \theref{theo:ber_1D_general.virtual}.


\bibliographystyle{IEEEtran}
\newpage
\bibliography{MyBibliography}
\end{document}